\documentclass{PoS}
\title{Charged Higgs in the NMSSM}

\ShortTitle{Charged Higgs in the NMSSM}
\author{A. G. Akeroyd\\
       Department of Physics, National Central University, 
Cheungli, Taiwan \\
        E-mail: \email{akeroyd@ncu.edu.tw }}
\author{\speaker{Abdesslam Arhrib}
\\
D\'epartement de Math\'ematiques, Facult\'e des Sciences et 
Techniques, B.P 416 Tangier, Morocco,\\
 LPHE, D\'epartement de Physique, Facult\'e des Sciences,
Marrakesh, Morocco   \\        E-mail: \email{aarhrib@ictp.it}}

\author{Qi-Shu Yan\\
   Physics Department,  University of Toronto, 60 St George Street, Toronto,\\
Ontario Canada M5S 117\\        E-mail: \email{yanqs.yan@utoronto.ca}}

\abstract{ 
The charged Higgs boson decays  $H^\pm\to W^\pm A_1$ and 
$H^\pm\to W^\pm h_i$ are studied in the 
framework of the next-to Minimal Supersymmetric Standard Model (NMSSM).
It is found that the decay rate for $H^\pm\to W^\pm A_1$ 
can dominate both below and above the top-bottom threshold. We suggest that  
$p p\to H^\pm A_1$ is a promising discovery channel for a
light charged Higgs boson in the NMSSM with small or moderate $\tan\beta$ 
and dominant decay mode $H^\pm \to W^\pm A_1$ which leads to $W^\pm A_1 A_1$.
This $W^\pm A_1 A_1$ signature can also arise from the Higgsstrahlung process
$pp\to W^\pm h_1$ followed by the decay $h_1\to A_1 A_1$.
It is shown that there exist regions of parameter space 
where these processes can have comparable cross sections and
we suggest that their respective signals can be distinguished 
at the LHC by using appropriate reconstruction methods. }

\FullConference{Prospects for Charged Higgs Discovery at Colliders\\
		 September 16-19 2008\\
		 Uppsala, Sweden}

\newcommand{\be}{\begin{equation}}
\newcommand{\ee}{\end{equation}}
\newcommand{\ba}{\begin{eqnarray}}
\newcommand{\beq}{\begin{equation}}
\newcommand{\eeq}{\end{equation}}
\newcommand{\ea}{\end{eqnarray}}
\def\bea{\begin{eqnarray}}
\def\eea{\end{eqnarray}}

\def\err#1#2{\lower2pt\hbox{ $\stackrel{\scriptstyle +#1}{\scriptstyle -#2}$}}
\def\ga{\mathrel{\raise.3ex\hbox{$>$\kern-.75em\lower1ex\hbox{$\sim$}}}}
\def\la{\mathrel{\raise.3ex\hbox{$<$\kern-.75em\lower1ex\hbox{$\sim$}}}}

\begin{document}

1. A charged Higgs boson ($H^\pm$) appears in any extension of the
Standard Model with two hypercharge Y=1 doublets. Its phenomenology has
been extensively studied in both the Two Higgs Doublet Model
(2HDM) and MSSM. The presence of $H^\pm$ is also predicted in the
 Next-to MSSM (NMSSM) in which an additional singlet neutral complex scalar
field $S$ is added to the two Higgs doublets of the MSSM.

In the NMSSM, after electroweak symmetry breaking
the Higgs spectrum  consists of three neutral 
scalars ($h_1$, $h_2$, $h_3$), two pseudoscalars ($A_1$, $A_2$) 
and a pair of charged Higgs bosons $H^\pm$. In both the CP-odd
and CP-even sector the physical eigenstates are ordered as
$M_{h_1}\la M_{h_2} \la M_{h_3}$ and $M_{A_1}\la M_{A_2}$.
For detailed discussions of the Higgs sector of the 
NMSSM the reader is referred to \cite{Drees:1988fc,Elliott:1993bs,
Franke:1995tc,Miller:2003ay,Accomando:2006ga}.
The mass of $H^\pm$ at tree-level is given by \cite{Drees:1988fc},
\cite{King:1995ys}:
\begin{equation}
M_{H^\pm}^2 =\frac{2 \mu_{eff}}{\sin 2\beta} (A_\lambda +\kappa s)+M_W^2
-\lambda^2 v^2 \label{chargedhiggs}
\end{equation}
where $\tan \beta =v_u/v_d$  and $v^2=v_u^2+v_d^2$.
This differs from the corresponding MSSM expression in which 
$M_A$ and $M_{H^\pm}$ are strongly correlated and become 
roughly equal for $M_A\geq 140$ GeV. \\
The CP-odd mass matrix can be obtained as follows:
Firstly, as in MSSM one rotates the bare 
fields $(\Im\rm{m}H_u, \Im\rm{m}H_d,\Im\rm{m}S)$
into a basis $(A,G,\Im\rm{m}S)$ where $G$ is a massless 
Goldstone boson.
Then one eliminates the Goldstone mode and the remaining 
$2\times 2$ CP-odd states are:
\begin{eqnarray}
A_1&=& \cos\theta_A A + \sin\theta_A \Im\rm{m}(S) \qquad , \qquad
A_2= -\sin\theta_1  A + \cos\theta_A \Im\rm{m}(S) 
\end{eqnarray}
Where $A=\cos\beta \Im\rm{m}H_u + \sin\beta \Im\rm{m}H_d$ 
is the CP-odd MSSM Higgs boson while $\Im\rm{m}S$ comes 
from the singlet $S$ field. 


In the MSSM the coupling $H^\pm A W$ (where $A$ is the
CP-odd neutral Higgs boson) contains
no mixing angle suppression but the relation $M_A\sim M_{H^\pm}$ 
ensures that the decay $H^{\pm} \to A W$ is greatly suppressed
in most of the parameter space. In the NMSSM, 
the relevant couplings for our study
are described by the following Lagrangian:
\begin{eqnarray}
{\cal L}_{VVH,VHH}& =&
gm_W g_{VVh_i}  W^{+\mu} W_{\mu}^- h_i -
g W_\mu^+(\frac{ig_{W^+H^-h_i}}{2} h_i  
+\frac{P_{i1}}{2} A_i)\stackrel{\leftrightarrow}{\partial}^\mu H^-
+ h.c
\label{gaugecoup}
\end{eqnarray}
where $g_{VVh_i}=\sin\beta S_{i1} +\cos\beta S_{i2}$, 
$g_{W^+H^-h_i}= \cos\beta S_{i1}-\sin\beta S_{i2}$, $P_{11}=\cos\theta_A$ 
and $P_{21}=-\sin\theta_A$, $S$ and $P$ are orthogonal matrices which 
diagonalize respectively the CP-even and CP-odd scalar mass matrix. 
From the last term in eq.~(\ref{gaugecoup}) one can see that the vertex
 $W^\pm H^\mp A_1$ is directly proportional to $P_{11}$
i.e. the doublet component of the mass eigenstate $A_1$. 
Consequently, if $A_1$ is entirely composed of doublet fields   
this coupling is maximized and if $A_1$ is purely singlet
the coupling vanishes.

Now we are ready to describe the phenomenology of the $H^\pm$
in the NMSSM and we summarize the results of our earlier work \cite{Akeroyd:2007yj}. 
The phenomenology of $H^\pm$ in the NMSSM 
has many similarities with that of $H^\pm$ in the MSSM.
This is to be expected since the fermionic couplings 
are identical in the two models.
The main differences in their phenomenology originate from the possibility 
of large mass splittings among the Higgs bosons in the NMSSM which permits 
decay channels like $H^{\pm} \to A_1W$ to proceed on-shell \cite{Drees:1998pw}. 
Moreover, in the NMSSM a light CP-even $h_1$ is also allowed and 
one can have the opening of the  decay $H^{\pm} \to h_1W$ both below and 
above the top-bottom threshold. This latter channel may change the NMSSM 
phenomenological predictions for
the charged Higgs with respect to the MSSM \cite{Drees:1998pw}.
In the MSSM the decay  $H^{\pm} \to h_1W$ is also open 
but the coupling $g_{W^+H^-h_1}\sim \cos^2(\beta-\alpha)$ is strongly
suppressed when $M_{H^\pm}\gg m_{h_1}+m_W$ and thus its branching
ratio is very small for such $M_{H^\pm}$.  For 
$M_{H^\pm}< m_{h_1}+m_W$ and just above the threshold the branching 
ratio for this channel can reach $10\%$ at most for small values 
of $\tan\beta$ \cite{Moretti:1994ds}, \cite{Djouadi:1995gv}, 
\cite{Drees:1999sb}.

The decay $H^\pm \to A W$, where $A$ is a CP-odd Higgs boson,
may be sizeable in a variety of models with a non-minimal Higgs sector
such as Two Higgs doublet models (Type I and II) 
\cite{Borzumati:1998xr,Akeroyd:1998dt,Akeroyd:2000xa} and in 
SUSY models with Higgs triplets \cite{DiazCruz:2007tf}.
Two LEP collaborations (OPAL and DELPHI) performed a search for 
a charged Higgs decaying to $A W^*$ (assuming $m_A> 2m_b$)
and derived limits on the charged Higgs mass
\cite{Abdallah:2003wd} comparable to those obtained from the
search for $H^\pm\to cs,\tau\nu$. In the MSSM the decay width 
for $H^\pm \to A W$ is very suppressed
in most of the parameter space \cite{Moretti:1994ds,Djouadi:1995gv}
because the charged Higgs and the CP-odd Higgs are close to mass
degeneracy.
The importance of the decays $H^\pm \to A_1 W$ and 
$H^\pm \to h_1 W$ in the NMSSM was first pointed out in \cite{Drees:1998pw}. 
Their branching ratios may be close to $100\%$
which can provide a clear signal at the LHC.

The decay width of $H^\pm \to A_1 W$ is directly proportional to 
$\cos\theta_A$ which is the doublet component
of $A_1$. This decay width can be substantially enhanced if 
$A_1$ is predominantly composed of doublet fields.
However, even with small doublet (large singlet) component of $A_1$ 
it is possible that $H^\pm \to A_1 W$ is the dominant decay mode. 
We perform a scan of the parameter space using the code 
NMSSM-Tools \cite{Ellwanger:2004xm} in order to quantify the importance
of $H^\pm \to A_1 W$ and $H^\pm \to h_1 W$.

Hereafter we assume that all scalar 
superparticles share the same soft mass term
$M_{SUSY}$, and the ratios of gaugino masses satisfy $M_1 : M_2 : M_3
= 1:2:6$; the trilinear couplings are related to $M_{SUSY}$ but
the sign is not fixed, {\it i.e.} $A_{t,b} = \pm 2 M_{SUSY}$. 
We scan the parameter space of the model by 
varying the free parameters within the following
region:
\bea
\lambda =[0, 1]\,, \,\,\,\,
\kappa =[-1, 1]\,,\,\,\,\,
\tan\beta = [0.2, 60]\,,\,\,\,\, 
\mu   =[-1, 1] \textrm{TeV} \,,\,\,\,\,  \nonumber \\
A_{\lambda}=[-1.0, 1.0] \textrm{TeV}\,,\,\,\,\,
A_{\kappa} =[-1.0,  1.0] \textrm{TeV}\,,\,\,\,\, 
M_{SUSY}   =[0.2, 3] \textrm{TeV}\,,\,\,\,\, 
M_{1}=[0.07, 3] \textrm{TeV}\,.
\eea
While varying these parameters, we take into account the experimental 
constraints on the MSSM spectrum e.g., charged Higgs mass $\ge 80$ GeV, 
chargino and scalar fermions $\ga 100$ GeV. We also apply the full set
of LEP constraints obtained from searches for neutral Higgs bosons
decaying to final states like $Z2b$, $Z4b$, $6b$, $6\tau$, $Z2b2\tau$,
$Z4\tau$, $2b2\tau$.

\begin{figure}
\begin{tabular}{cc}
\resizebox{82mm}{!}{\includegraphics{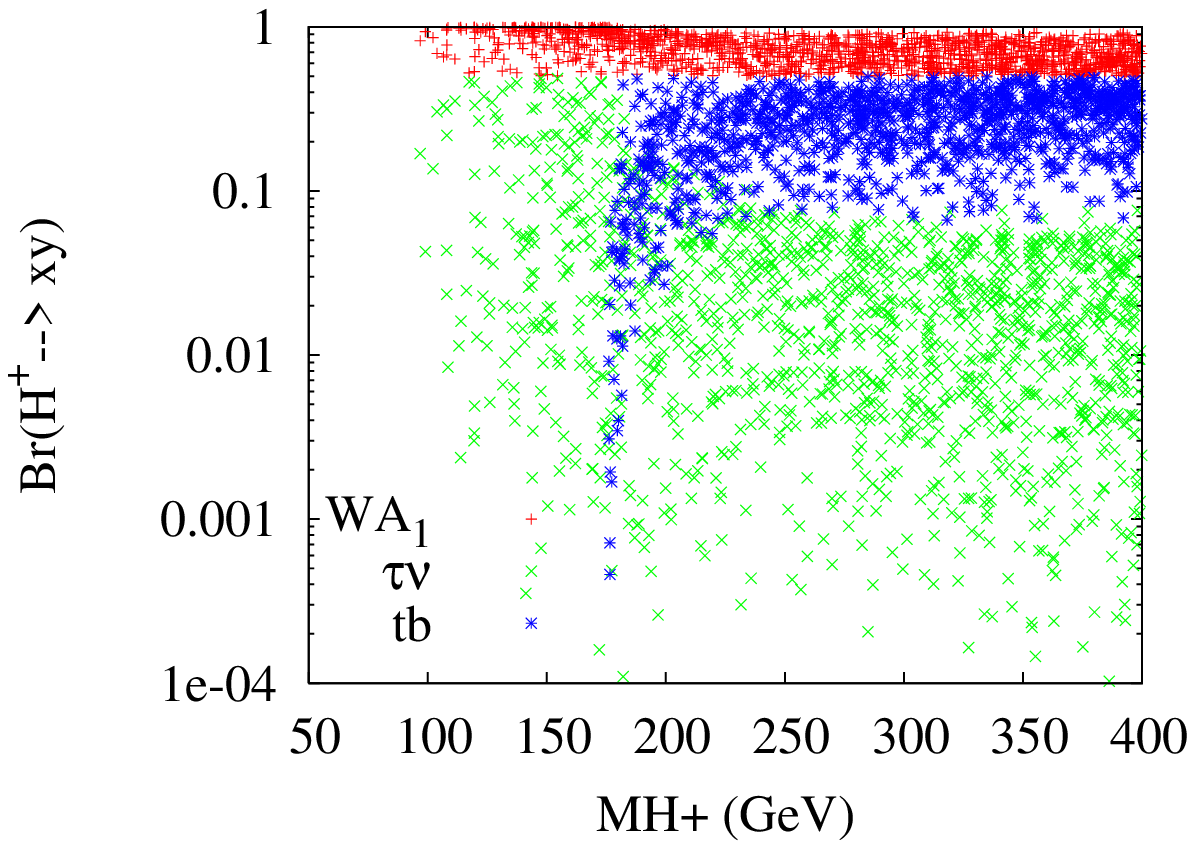}} & \hspace{-1.6cm}
\resizebox{82mm}{!}{\includegraphics{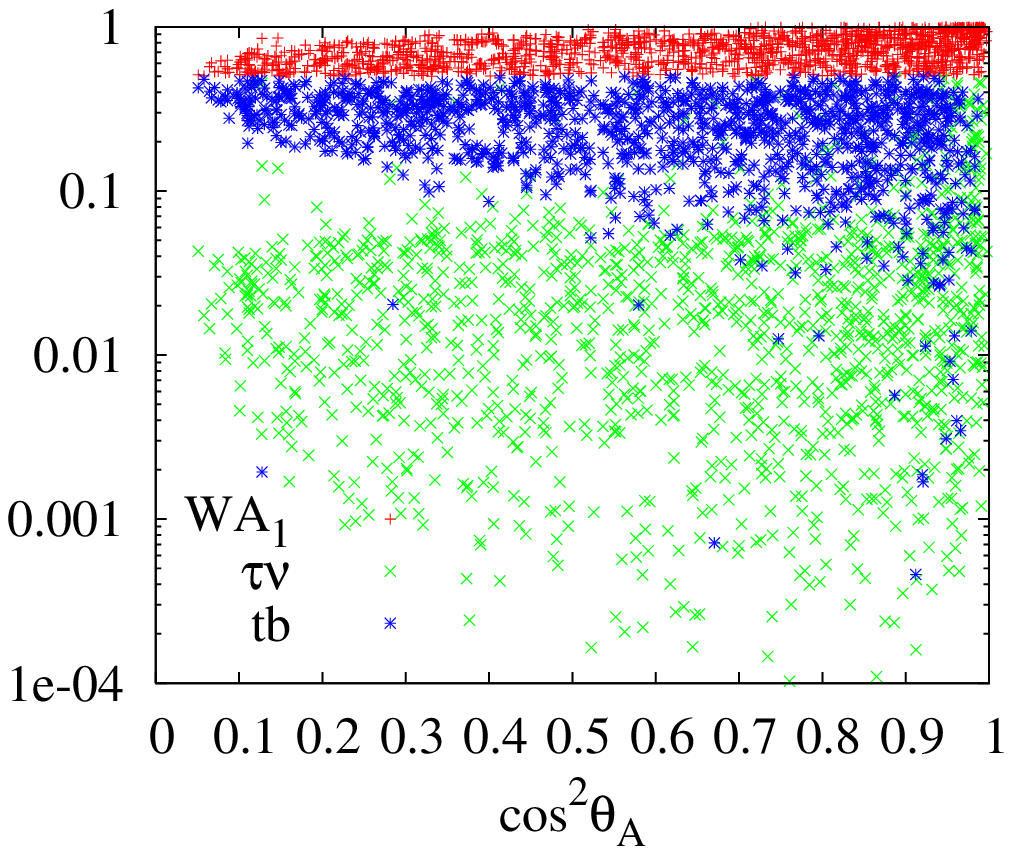}}\\
\end{tabular}
\caption
{\it Comparison of the branching ratios of $H^\pm \to \{W^\pm A_1,\tau\nu ,tb\}$
as a function of $M_{H^\pm}$ (left), $\cos\theta_A$ (right).
In all panels only points with $Br(H^\pm \to  W^\pm A_1)\geq 50 \%$
are selected.}
\label{plot1}
\end{figure}

In Fig.~(\ref{plot1}) we display the branching ratios of 
$W^\pm A_1$ , $\tau\nu$  and top-bottom modes.
Before the opening of the $H^\pm \to tb$ channel,
the full dominance of $W^\pm A_1$ over $\tau\nu$ requires
light $M_{A_1} \la 100 \,GeV$, large doublet component of $A_1$ 
and $\tan\beta$ not too large. Note that at large 
$\tan\beta \approx 15-25$, the $W^\pm A_1$ and $\tau\nu$ 
channels become comparable in size.
Once the decay $H^\pm \to tb$ is open, it competes strongly with 
$W^\pm A_1$ for $\tan\beta\la 15$. As can be seen from 
Fig.~(\ref{plot1}) left, the branching ratio of 
$H^\pm\to W^\pm A_1$ is less than $90\%$. It is interesting
to see also that for $\cos^2\theta_A \la 0.05$ there is not a 
single point with $Br(H^\pm\to W^\pm A_1)\ga 50\% $.
Note also that at large $\tan\beta\ga 25$, it is hard for 
$H^\pm\to W^\pm A_1$ to compete with $\tau\nu$  and top-bottom modes.


The most problematic region for $H^\pm$ discovery in the MSSM
is for moderate values 
of $\tan\beta$, since the production mechanisms which rely on a large 
bottom quark or top quark Yukawa coupling (e.g. $gb \to H^\pm t$)
are least effective.
Hence alternative mechanisms which could offer good detection 
prospects for $H^\pm$ at moderate values of $\tan\beta$ are desirable.
The cross sections for the
pair production mechanisms $pp \rightarrow H^\pm A_1$ and
$pp \rightarrow H^\pm h_1$ fall quickly with increasing
scalar masses but for relatively light masses 
($\la 200$ GeV) they can provide promising signal rates
which might enable their detection at the LHC (see \cite{Kanemura:2001hz} for
studies in the context of the MSSM). 
One common feature is that the produced scalars
enjoy large transverse momenta, which are crucial for the 
trigger and event selection. 

In the NMSSM, if the coupling $H^\pm W^\mp A_1$ is  sizeable,
so will be the cross section for $pp\to W^\pm \to H^\pm A_1$
provided that $H^\pm$ and  $A_1$ are not too heavy.
The production mechanism $pp\to H^\pm A_1$ 
followed by the decay $H^\pm\to W^\pm A_1$  
would give rise to a signal $W^\pm A_1A_1\to Wbbbb$ 
\cite{Akeroyd:2003jp} or $W^\pm A_1A_1\to W\tau\tau\tau\tau$.
The signature $W^\pm A_1A_1\to Wbbbb$ was simulated at the 
LHC in \cite{Ghosh:2004wr}
in the context of the CP violating MSSM with the conclusion that a 
sizeable signal essentially free of background could be obtained.
We use NMSSM-TOOLS1.1.1 to calculate the mass spectrum and 
couplings of the NMSSM
Higgs bosons, and we link CTQ6.1M PDF distribution to this code 
in order to calculate the cross sections of $pp \rightarrow H^\pm A_1$,
$pp \rightarrow H^\pm h_1$ and $pp \rightarrow W^\pm h_1$.
All cross sections are evaluated at a scale which is the sum of the masses
in the final states and do not include
next-to-leading order QCD enhancement factors (K factors) of around $1.2
\to 1.3$ \cite{Kanemura:2001hz},\cite{Han:1991ia}.



Note that the process $pp\to H^\pm A_1\to W^\pm A_1A_1$ leads to the same
signature as the process $pp \to Wh_1 \to WA_1A_1\to Wbbbb$.
The latter has been simulated in \cite{Cheung:2007sv} and also offers
very good detection prospects. We will compare the magnitude of these two
distinct mechanisms which lead to the same $Wbbbb$ signature. In addition, 
the mechanism $pp\to H^\pm h_{1}$ followed by the decay $H^\pm\to W^\pm A_1$ 
would also lead to the same final state $W^\pm A_1h_1\to Wbbbb$.

\begin{figure}
\begin{tabular}{cc}
\hspace{-.5cm}\resizebox{83mm}{!}{\includegraphics{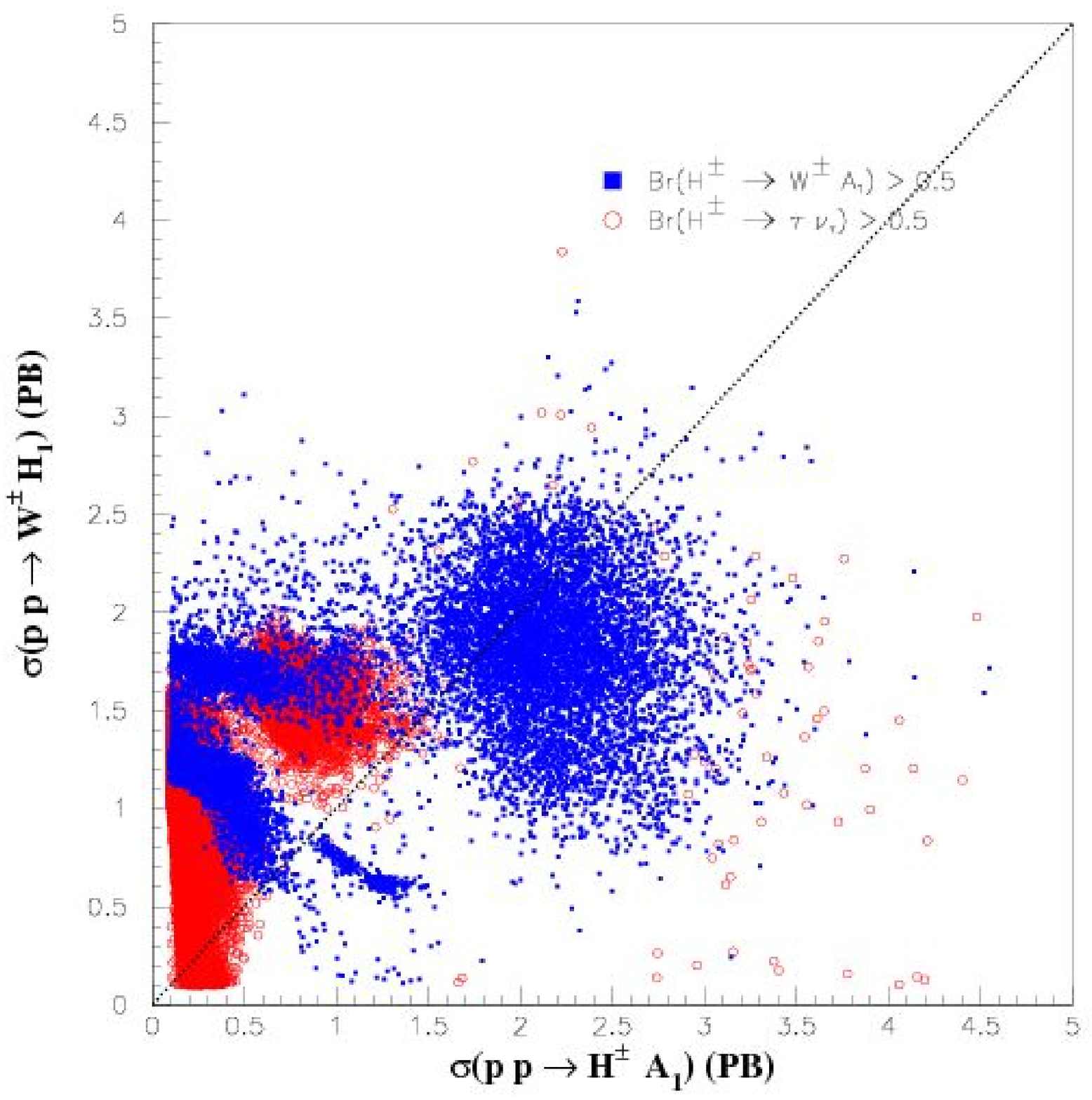}} & \hspace{-1.35cm}
\resizebox{83mm}{!}{\includegraphics{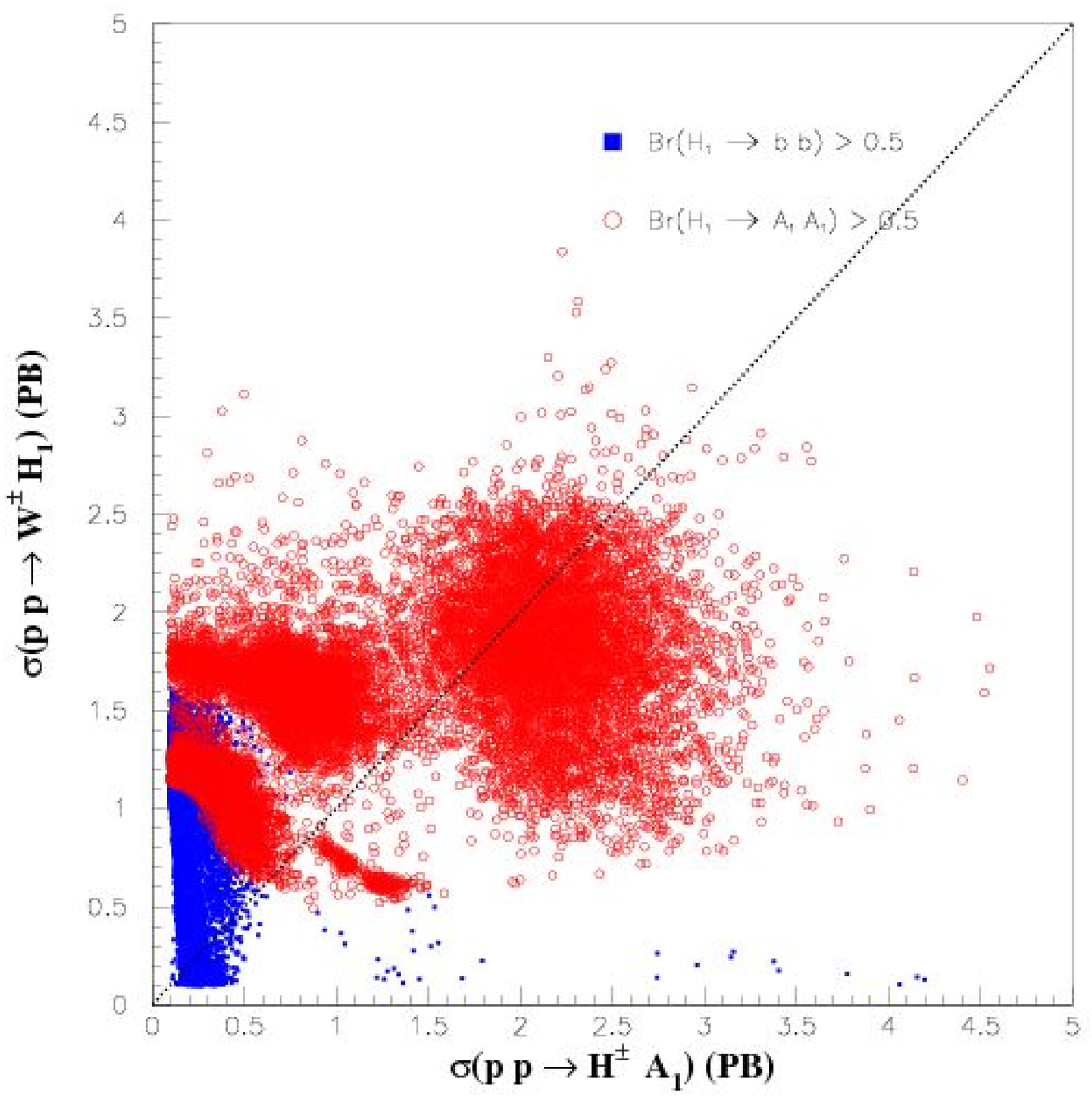}}
\end{tabular}
\caption{\it{ Left panel: comparison of $\sigma(pp \rightarrow H^\pm A_1)$ 
and $\sigma(pp \rightarrow W^\pm h_1) $ with two $H^\pm$ decay 
modes.  Right panel: comparison of $\sigma(pp \rightarrow H^\pm A_1)$ 
and $\sigma(pp \rightarrow W^\pm h_1) $ with two $h_1$ decay 
modes. The dotted line corresponds to
$\sigma(pp \rightarrow W^\pm h_1) = \sigma(pp \rightarrow H^\pm A_1)$. }}
\label{fig3}
\end{figure}

Hence a numerical comparison of their cross sections
is of particular interest and is shown in Fig.~(\ref{fig3}), where
all points satisfy the following conditions:
\begin{eqnarray}
\sigma(pp \rightarrow H^\pm A_1) > 0.1 \ pb\, \qquad \rm{and} \qquad
\sigma(pp \rightarrow W^\pm h_1) > 0.1  \ pb \,.
\label{cond-vvh-ca}
\end{eqnarray}
Superimposed on Fig.~(\ref{fig3}a) and Fig.~(\ref{fig3}b) are the main decay 
modes of the charged Higgs boson and the decay neutral Higgs boson $H_1$ 
respectively. We further impose the following conditions:
\begin{eqnarray}
Br(H^\pm \to W^\pm A_1) > 0.5 \qquad \rm{and} \qquad
Br(h_1 \to A_1 A_1) > 0.5 \,,
\label{conditions}
\end{eqnarray}
and the surviving points are displayed in Fig. (\ref{fig4}a). Importantly, 
there are many points where the two cross sections are of comparable size. 
We note that for these points in Fig. (\ref{fig4}a)  the pseudoscalar $A_1$ 
can be both R-axion like or a mixture of the three allowed basic axions.
If the magnitude of the cross sections of both $pp \rightarrow H^\pm A_1$ and 
$pp \rightarrow V h_1$ are similar then the interference of the two channels 
(i.e., the same $Wbbbb$ signature arising from distinct production mechanisms)
should be taken into account. We have neglected such effects in the present 
study.

We now discuss whether the $Wbbbb$ signatures can be distinguished 
experimentally by comparing the strategies adopted in 
\cite{Ghosh:2004wr} (for $pp\to H^\pm A^0$)
and \cite{Cheung:2007sv} (for $pp\to W^\pm h_1$).
In order to reconstruct the peak of the CP-even Higgs $h_1$,
one can select events with a charged lepton and four tagged $b$ quark jets 
as shown in \cite{Cheung:2007sv}.
This enables both a clean Higgs signal 
with high significance and a measurement of
$M_{h_1}$ given by the invariant mass of the four $b$ quark jets,
$m_{4b}$. The process $pp\to H^\pm A_1$ might be an irreducible
background but presumably could be significantly 
suppressed with the aforementioned cut on $m_{4b}$
e.g., $m_{h_1}-15 {\rm GeV} < m_{4b} < m_{h_1}+15 {\rm GeV}$.

Regarding detection of $pp\to H^\pm A^0$, it was demonstrated 
in~\cite{Ghosh:2004wr}
(for the analogous process $pp\to H^\pm H_1\to W H_1 H_1$ in the 
CP violating MSSM) that the mass of $H^\pm$ can be reconstructed.
This is achieved by defining a tranverse mass ($M_T$) 
which is a function of the momenta of the two secondary $b$ jets 
(i.e., those originating from the decay $H^\pm\to A_1W\to Wbb$) 
and the momenta of the lepton and missing energy 
coming from the $W$ boson. It was shown that
$M_T$ is sensitive to the underlying charged Higgs mass and 
thus can be used for the determination of $M_{H^\pm}$.
The pair of $b$ jets from $pp\to W^\pm h_1$ might be an irreducible
background but presumably could be suppressed with a cut on $M_T$

To reconstruct the peak of the light CP-odd neutral
Higgs $A_1$ one can require events with
four tagged $b$ jets, construct the three possible double pairings of 
$b\bar{b}$ invariant masses, and then select the pairing giving the least
difference between the two $b\bar{b}$ invariant masses values
\cite{Ghosh:2004wr}.
$W 4b$ signatures from the process $pp\to W^\pm h_1$ also 
contribute constructively to the reconstruction of $A_1$.
Thus we conclude that it is promising to reconstruct the peaks of 
the CP-even neutral Higgs ($h_1$), charged Higgs ($H^\pm$) 
and CP-odd neutral Higgs ($A_1$) and thus 
experimentally distinguish the $Wbbbb$ signatures arising 
from the two distinct production mechanisms. 
We defer a detailed simulation to a future work.

In summary, It was shown that $H^\pm\to W^\pm A_1$ 
can dominate over the standard
decays $H^\pm\to \tau^\pm\nu$ and $H^\pm\to tb$ 
both below and above the top-bottom threshold. 
Large branching ratios for $H^\pm\to W^\pm A_1$ and 
$H^\pm\to W^\pm h_1$ would affect
the anticipated search potential for $H^\pm$ at the LHC.
We also studied the production process
$pp\to H^\pm A_1$ and showed that sizeable cross sections
($> 1$ pb) are possible. 
It is known that intermediate values of $\tan\beta$ 
(e.g., $5 < \tan\beta < 20$) are
most problematic for discovery of $H^\pm$ at the LHC \cite{Assamagan:2002ne}
since the $H^\pm tb$ Yukawa coupling (which is employed in the 
conventional production processes) takes its lowest values.
In such a region the process $pp \to H^\pm A_1$
can have a sizeable cross section if $m_{H^\pm}+m_{A_1} < 200$ GeV.
Therefore we propose $pp \to H^\pm A_1$ as a unique mechanism
to probe the parameter space of intermediate $\tan\beta$ and light charged 
Higgs boson in the NMSSM.

\section*{Acknowledgments} 
A.A thanks the organiser for the warm hospitality extended to him during 
the workshop.

\end{document}